\documentclass[manuscript]{aastex}

\newcommand{\markRevision}[1]{{#1}}

\shorttitle{Distribution of Discontinuities in 3D MHD Simulation}
\shortauthors{Zhang et al.}

\begin{document}

\title{Occurrence Rates and Heating Effects of Tangential and Rotational
Discontinuities as Obtained from Three-dimensional Simulation of
Magnetohydrodynamic Turbulence}

\author{Lei Zhang, Jiansen He, Chuanyi Tu}
\email{jshept@pku.edu.cn}
\affil{School of Earth and Space Sciences, Peking University, Beijing, China, 100871}

\author{Liping Yang}
\affil{%
SIGMA Weather Group, State Key Laboratory for Space Weather, Center for
Space Science and Applied Research, Chinese Academy of Sciences, Beijing,
China
\\
School of Earth and Space Sciences, Peking University, Beijing, China, 100871}

\author{Xin Wang}
\affil{School of Earth and Space Sciences, Peking University, Beijing, China, 100871}

\author{Eckart Marsch}
\affil{Institute for Experimental and Applied Physics, Christian-Albrechts-Universit\"at zu Kiel, Germany}

\author{Linghua Wang}
\affil{School of Earth and Space Sciences, Peking University, Beijing, China, 100871}
\begin{abstract}

In solar wind, magnetohydrodynamic (MHD) discontinuities are ubiquitous and
often found to be at the origin of turbulence intermittency. They may also
play a key role in the turbulence dissipation and heating of the solar wind.
The tangential (TD) and rotational (RD) discontinuities are the two most
important types of discontinuities. Recently, the connection between
turbulence intermittency and proton thermodynamics has been being
investigated observationally. Here we present numerical results from
three-dimensional MHD simulation \markRevision{with pressure anisotropy} and define new methods to identify and to
distinguish TDs and RDs. Three statistical results obtained about the
relative occurrence rates and heating effects are highlighted:
  (1) RDs tend to take up the majority of the discontinuities along with time;
  (2) the thermal states embedding TDs tend to be associated with extreme plasma
  parameters or instabilities, while RDs do not;
  (3) TDs have a higher average $T$ as well as perpendicular temperature $T_\perp$.
The simulation shows that TDs and RDs evolve and contribute to solar wind
heating differently. These results will inspire our understanding of the
mechanisms that generate discontinuities and cause plasma heating.
\end{abstract}


\section{Introduction}

The turbulent solar wind embodies discontinuities \citep[e.g.~][]{colburn66,
tu95, marsch06, bruno13, paschmann13}. The tangential discontinuity (TD) and
rotational discontinuity (RD) are the two most important yet quite different
types: in the deHoffmann-Teller frame, theoretically, TDs have no normal
components of $\mathbf{v}$ and $\mathbf{B}$ on both sides, while RDs have
normal components, must obey the Wal\'en relation $\mathbf{v}=\pm \mathbf{B}/\sqrt{\mu_0\rho}$, 
and keep $\left|\mathbf{v}\right|$ and $\left|\mathbf{B}\right|$ continuous. Furthermore, TDs allow jumps of density and
temperature, while these parameters have the same values on both sides of the
RDs. As to the mechanisms of how discontinuities form, there exist two kinds
of explanations. There are empirical evidences indicating that
discontinuities are boundaries of magnetic flux tubes \citep{borovsky08,
miao11}, and there are suppositions that they form locally through nonlinear
interactions and may be associated with small random currents
\citep{greco09}. In \cite{servidio11}'s 2D magnetohydrodynamic (MHD)
simulation, most discontinuities appear to be TDs. However, in a 3D geometry,
it remains unknown which type of discontinuity dominates.

Dissipation of turbulence is considered an important contributor to the
heating of the solar wind. Many recent studies concentrated on the role of
intermittency and discontinuities in this process. \cite{bale09} discovered
strongly enhanced fluctuations along the thresholds of plasma instabilities.
\cite{osman11} reported that high PVI (Partial Variance Increment) levels of
various parameters correspond to intensive plasma heating and higher
temperatures of electrons as well as ions. \cite{osman12} researched a large
sample of data from measurements made by the \textit{Wind} spacecraft and
plotted the data distributions in the $(\beta_\parallel,A)$ parameter plane
($\beta_\parallel=p_{\parallel{p}}/(B^2/(2\mu_0))$, $A=T_\perp/T_\parallel$). Thus they could show that the distributions are
roughly \markRevision{bounded} by curves corresponding to the mirror and oblique fire-hose
instabilities, that the regions near the instability thresholds have higher
averaged PVI, and that events with intense PVIs have $A$ far from unity.
\cite{wang13} analysed observed discontinuities with the PVI technique and
found that the majority of them are RDs, but TDs have more obvious proton
temperature increases. These empirical findings inspired us to investigate
also the heating effects at the TDs and RDs obtained in our simulation.

Numerical simulations have been employed before to understand plasma heating.
\cite{greco08} assigned a path through the computational domain and then
adopted the notion of PVI to analyse the 3D simulation data sampled along
that path. Within their Hall MHD model they thus identified small-scale
discontinuities being associated with intermittency. \cite{parashar09}
demonstrated by use of a 2.5D hybrid model that an ion temperature anisotropy
can be created while the protons are heated by magnetic energy dissipation.
\cite{karimabadi13} conducted a full particle simulation which showed that,
triggered by the Kelvin-Helmholtz instability with strong velocity shear, a
turbulent cascade generates current sheets heating the plasma locally, and
which yielded anisotropic particle distributions in that process.
\cite{servidio14} allowed for a broader range of $\beta_\parallel$ and the
strength of the magnetic field fluctuations, thus obtaining results that
basically are in accordance with those of \cite{osman12}. However, neither
were the data with high PVIs investigated, nor was the related heating of
particles studied. All the mentioned work did also not distinguish between
the various types of discontinuities in their simulation data.

Motivated by all these aspects, we will here conduct a 3D numerical
simulation with the aim to test new numerical methods to identify and analyse
discontinuities, without assuming auxiliary paths along which data are
sampled in the simulation domain. Based on this discontinuity identification,
we present statistical results in order to investigate the proportion of TDs
and RDs in all the discontinuities found, their parameter distribution in the
$(\beta_\parallel,A)$ plane, and the temperature increases in TDs and RDs.
Such counts of TDs and RDs and their distributions have, to our knowledge,
not been reported previously. These new simulation results will help us to
understand better particle heating at intermittent structures in the solar
wind, and thus to resolve the turbulence dissipation problem.

In Section 2 we will describe the numerical tools and the methods employed to
identify and categorize discontinuities. In Section 3 we present a specific
case of a TD and and RD, as well as statistical properties of all
discontinuities found in the computation domain. In Section 4 we shall
summarize our study, and further discuss relevant physical issues.

\section{Methods}

\subsection{Numerical Model of MHD Turbulence}

In order to evaluate the role of temperature anisotropy in MHD turbulence, we
adopt the model which employs the compressible ideal MHD equations and
incorporates an anisotropic pressure tensor \citep{meng12, meng12b, meng13}:
\begin{equation}
  \frac{\partial\rho}{\partial{t}}+\nabla\cdot(\rho\mathbf{v})=0,
\end{equation}
\begin{equation}
  \frac{\partial\rho\mathbf{v}}{\partial{t}}+\nabla\cdot\left(\rho\mathbf{v}\mathbf{v}+p_\perp\mathsf{I}+(p_\parallel-p_\perp)\hat{\mathbf{B}}\hat{\mathbf{B}}-(\mathbf{B}\mathbf{B}-B^2\mathsf{I}/2)/\mu_0\right)=0,
\end{equation}
\begin{equation}
  \frac{\partial\mathbf{B}}{\partial{t}}+\nabla\times(-\mathbf{u}\times\mathbf{B})=0,
  \label{eqn:magFreeze}
\end{equation}
\begin{equation}
  \frac{\partial{p}_\parallel}{\partial{t}}+\nabla\cdot(p_\parallel\mathbf{u})+2p_\parallel\hat{\mathbf{B}}\cdot(\hat{\mathbf{B}}\cdot\nabla)\mathbf{u}=\frac{\delta{p}_\parallel}{\delta{t}},
  \label{eqn:pPara}
\end{equation}
\begin{equation}
  \frac{\partial{p}}{\partial{t}}+\nabla\cdot(p\mathbf{u})+2p_\perp\nabla\cdot\mathbf{u}/3+2(p_\parallel-p_\perp)\hat{\mathbf{B}}\cdot(\hat{\mathbf{B}}\cdot\nabla)\mathbf{u}/3=0,
  \label{eqn:pTotal}
\end{equation}
where $p=(p_\parallel+2p_\perp)/3$, $\hat{\mathbf{B}}=\mathbf{B}/\left|B\right|$. To describe instabilities correctly, it is common to use a
heuristic term of pressure relaxation restricting the temperature anisotropy
\citep{hesse92, birn95} denoted $\delta{p_\parallel}/\delta{t}$.
\markRevision{In our numerical simulation we employ this modified MHD model, which is not self-consistent as 
Vlasov models \citep[e.g.][]{servidio12, servidio14}. 
This simplification of our model, on the other hand, 
will help to understand the role of fluid-like closures of the pressure
tensor, which would be dissipative with the pressure-transport terms. 
}
\markRevision{Equations~\ref{eqn:pPara} and \ref{eqn:pTotal} keep 
the quantities $p_\perp/(nB)$ and $B^2p_\parallel/n^3$
adiabatically invariant \citep{cgl}, 
if the RHS of Equation~\ref{eqn:pPara} is zero. 
As Equation~\ref{eqn:magFreeze} is of the ideal MHD type, 
magnetic helicity is conserved.
However, the model still lacks important kinetic physics of the solar wind, 
e.g. Landau damping and ion cyclotron resonances, 
which can be vital to turbulence dissipation. }

To solve the above system of equations, we utilize the BATSRUS codes
\citep{powell99, toth12}. The simulation is conducted in a three-dimensional
cartesian coordinate system and encompasses an absolute volume of
$(62.8\;\textrm{Mm})^3$ that is resolved in $256^3$ grid points. 
\markRevision{The grid resolution ($\sim250$~km) is well above the ion skin 
depth ($\sim100$~km), so Hall-dispersive physics is not included. }
We use the scheme proposed by \cite{rempel09} in order to control numerical diffusion,
and by applying such diffusion control strictly keep $\nabla\cdot\mathbf{B}=0$. 
This method guarantees proper dissipation and correct jump conditions
at discontinuities. 
\markRevision{No explicit diffusive term is included in the numerical 
simulation code. 
Yet the magnetic and kinetic energy are subject to decay numerically. 
This decay can be attributed to the numerical scheme and grid resolution. 
}
We apply periodic boundary conditions to all the six
surfaces of the simulation box. The initial conditions are set uniformly for
$\rho$, $T_\parallel$ and $T_\perp$, and randomly for $\mathbf{v}$ and
$\mathbf{B}$, with the average $\mathbf{v}_0=0$ and a finite guide field
$\mathbf{B}_0\parallel\hat{\mathbf{e}}_z$. To simulate the solar wind at 1~AU, the
field $\mathbf{B}_0$ is chosen as 5~nT, while the Alfv\'en speed is set at
50~km~s$^{-1}$, and $p=p_\parallel=p_\perp$ with
$\sqrt{5p/(3\rho)}=50\;\mbox{km}\;\mbox{s}^{-1}$, which corresponds to a proton
number density of 5~cm$^{-3}$ and an isotropic temperature of $10^5\;\mathrm{K}$. 
For the turbulence part of the fields we take Fourier amplitudes obeying
the broadband initial conditions described by \cite{matthaeus96}, with
$\left|\delta\mathbf{B}(\mathbf{k})\right|^2=\left|\delta\mathbf{v}(\mathbf{k})\right|^2\propto(1+k/k_\textrm{knee})^{-q}$ in the range
$10^{-7}\;\mathrm m^{-1}\le{k}\le8\times10^{-7}\;\mathrm{m}^{-1}$. The
parameter $k_\textrm{knee}$ is set to be $3\times10^{-7}\;\mathrm{m}^{-1}$,
and the spectral index $q$ is set so that the slopes of the power spectral
densities are both $-5/3$. We take $\sqrt{\left<v^2\right>}=30\;\textrm{km s}^{-1}$ 
(accordingly $\sqrt{\left<\delta{B}^2\right>}=3\;\textrm{nT}$). 
The dimensionless cross-helicity $\sigma=0.9$.

\subsection{Methods of Selecting and Categorizing Discontinuities}

The analysis of discontinuities calls for reliable methods to select and
categorize them in observational or numerical data. To trace abrupt spatial
changes of the magnetic field, we use the total variance of increments (TVI)
as an indicator and set a threshold for it. To calculate TVI, the total
variance is first computed as
\begin{equation}
  \left|\Delta{B}\right|=\sqrt{\sum_{\alpha,\beta=x,y,z}(\partial{B}_\beta/\partial\alpha)^2},
\end{equation}
where the partial derivative at grid point $(i,j,k)$ about $x$ is computed
as
\begin{equation}
  \frac{B_\bullet(i+w,j,k)-B_\bullet(i-w,j,k)}{2w\delta{x}}.
\end{equation}
Here $\delta{x}$ is the grid distance, $B_\bullet$ denotes the corresponding
component of magnetic field, and $w$ is the width (in this work, we take $w=3$, i.e.~within a cuboid of $7^3$ grid points). 
For the $y$ and $z$
derivatives, similar differences along the corresponding directions are used.
Then the TVI is the normalized total variance
\begin{equation}
  \textrm{TVI}=\left|\Delta{B}\right|/\left<\left|\Delta{B}\right|\right>,
\end{equation}
where the angle brackets denote the average over the whole computational
domain. This definition is a further development of the previous PVI defined
by \cite{greco08} (similar to the method adopted before by \cite{marsch94}),
which was taken along a given path and hence directionally sensitive. 
Yet the above TVI includes all directions and thus is unbiased for all points
possibly belonging to a discontinuity. 
\markRevision{The TVI utilises more information from fully 3D data, 
and accordingly gives more physical insight. 
However, most solar wind measurements are 1D samples, so the method is inapplicable to such measurements. }
In the present work, those grid points
with $\textrm{TVI}>3$ are actually identifiable as discontinuity points and
chosen for subsequent analysis.

At each discontinuity point, we conduct the minimal variance analysis (MVA)
\citep{sonnerup67} in its neighbourhood, defined as a cuboid of the same
given size for all points considered. Since $B_{n1}=B_{n2}$ is required,
the direction of minimal variation can be regarded as the normal of the
discontinuity \citep{sonnerup67}. 
In this work, we just consider a discontinuity locally as a small plane that contains
the discontinuity point and cuts its neighbouring cuboid into two segments,
so that averages of quantities on either side of the plane can be calculated
in the segments obtained.

To categorize the discontinuities, we use the criteria defined by
\cite{smith73}, which aim at judging two features: (1) whether $B_n=0$
holds, and (2) whether $\left|B\right|$ remains continuous. Hence the
parameters $P_1=\left|B_n\right|/B_L$ and $P_2=\delta\left|B\right|/B_L$ are calculated ($B_L$ is the larger of $\left<\left|B\right|\right>$ on
both sides). The points with $P_1<0.2$ and $P_2>0.2$ are categorized as
TD, while the ones with $P_1>0.4$ and $P_2<0.2$ as RD\@. All the other
cases are categorized as ED(\markRevision{either TD or RD type}, with $P_1<0.4$ and 
$P_2<0.2$) and ND(\markRevision{neither TD nor RD type}, with $P_1>0.2$ and $P_2>0.2$).

The aforementioned analysis only involves magnetic field data. To check and
corroborate the results thus obtained, we also analyse the plasma velocity,
density, and temperature data. In such case studies, it is trivial to check
whether the density or temperature is continuous, but the velocity has to
undergo the Wal\'en test for RDs, i.e. one has to test whether 
$\mathbf{v}-\mathbf{v}_\textrm{HT}=\pm\mathbf{b}=\pm\mathbf{B}/\sqrt{\mu_0\rho}$,
which is usually done statistically in a scatter plot. 
The Wal\'en test must be conducted in the deHoffmann-Teller frame. 
To find its velocity $\mathbf{v}_\textrm{HT}$, 
in the cuboid the sum $\sum_{(i,j,k)}\left|(\mathbf{v}_{ijk}-\mathbf{v}_0)\times\mathbf{B}_{ijk}\right|^2$ is to be minimized.
The $\mathbf{v}_0$ that makes this sum a minimum is then accepted as velocity
of the deHoffmann-Teller frame \citep{sonnerup87}.

\section{Results}

The methods described above permit us to select the TDs and RDs from all
cases with large TVI, and so we obtain corresponding data sets of
discontinuities on which we can do individual and statistical research. 

\begin{figure}
  \plotone{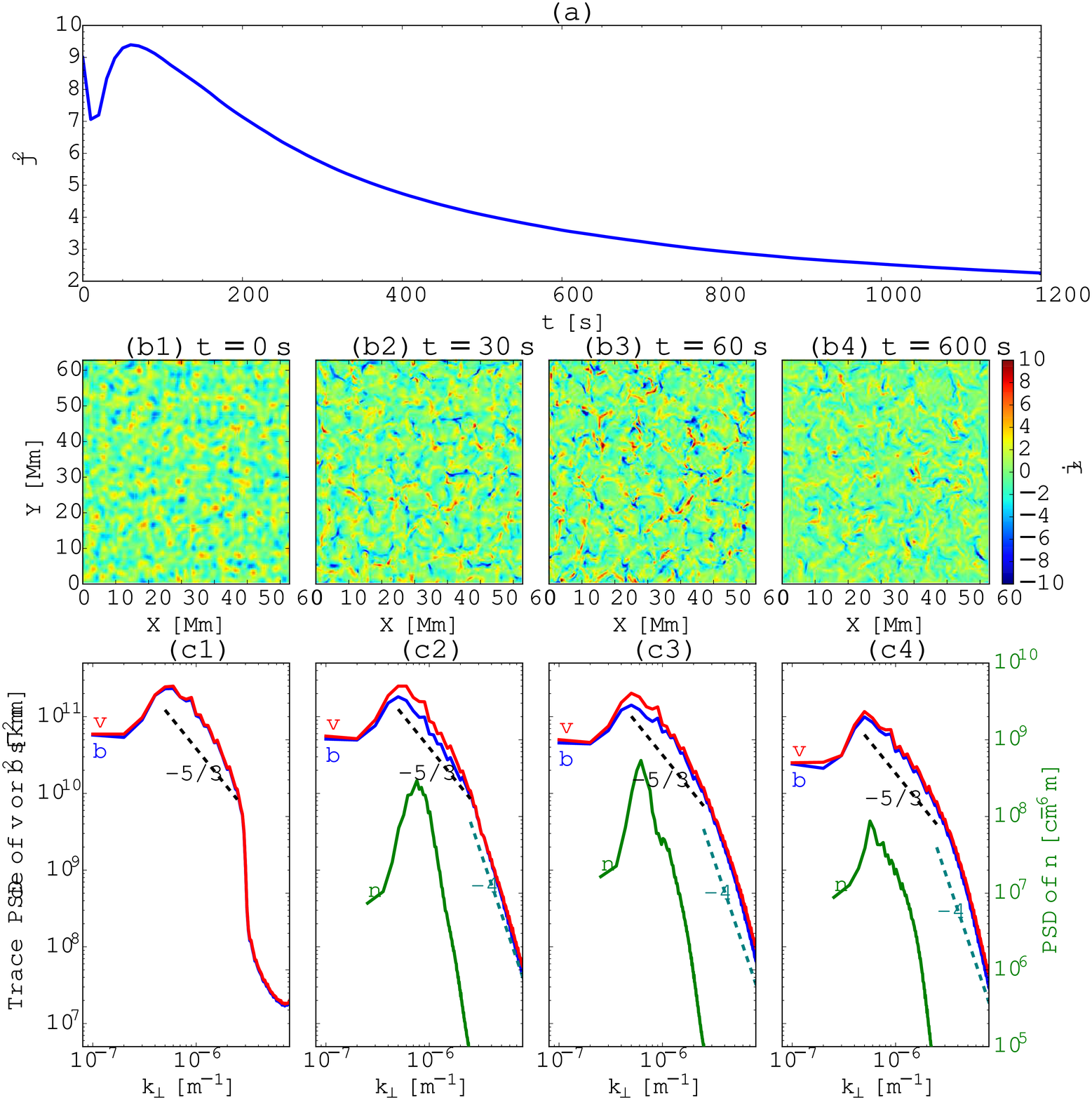}
  \caption{%
  Evolution stages of the decaying turbulence in the MHD simulation.
  \textbf{(a)}: Temporal profile of $\left<j^2\right>$ in arbitrary unit.
  \textbf{(b)}: Distribution of the $z$-component of the current density $j_z$ in
  the plane $z=31.4\;\mathrm{Mm}$ at the given times; 
  $j_z$ is given in arbitrary unit in agreement with (a).
  \markRevision{\textbf{(c)}: Trace power spectral density of $\mathbf{v}$ (red) and
  $\mathbf{b}$ (blue), as well as PSD of $n$ (green) at the same times. 
  The horizontal axis represents $k_\perp=\sqrt{k_x^2+k_y^2}$, while the
  vertical axis represents $\mathbf{v}$ or $\mathbf{b}$(black, left) and $n$(green, right). 
  Spectral indices $-5/3$(black) and $-4$(cyan) are also plotted in dashed
  lines for reference.}\label{fig1}}
\end{figure}

To understand the evolution of the decaying turbulence in our MHD simulation,
we plot in Figure~\ref{fig1} the squared current density $\left<j^2\right>$
averaged over the whole computational domain. This quantity relates to the
curl of the magnetic field and describes its inhomogeneity and energy
conversion (dissipation). Its evolution in time clearly shows the following
phases: an initial drop, subsequent increase and final decay. This whole
trend basically agrees with that found by \cite{matthaeus96} in their
previous simulation. To further illustrate that process, the $z$-component of current density $j_z$ is also plotted
in a space-time display. 
The evolution implies that the initial drop of
$\left<j^2\right>$ is due to a consumption of the magnetic energy in the
compressive turbulent plasma motion, which leads to growing density
inhomogeneity. This increase involves the formation of thin and stretched
current sheets (see the numerous thin and sharp structures in (b3)).
Due to the decay process, the inhomogeneities in (b4) fade
away, yet a few current-sheet-like structures remain.
\markRevision{The trace power spectral density of $\mathbf{v}$, $\mathbf{b}$, 
and power spectral density of $n$ are also shown. 
A region with spectral index close to $-5/3$ seems to be identified. }

\begin{figure}
  \plotone{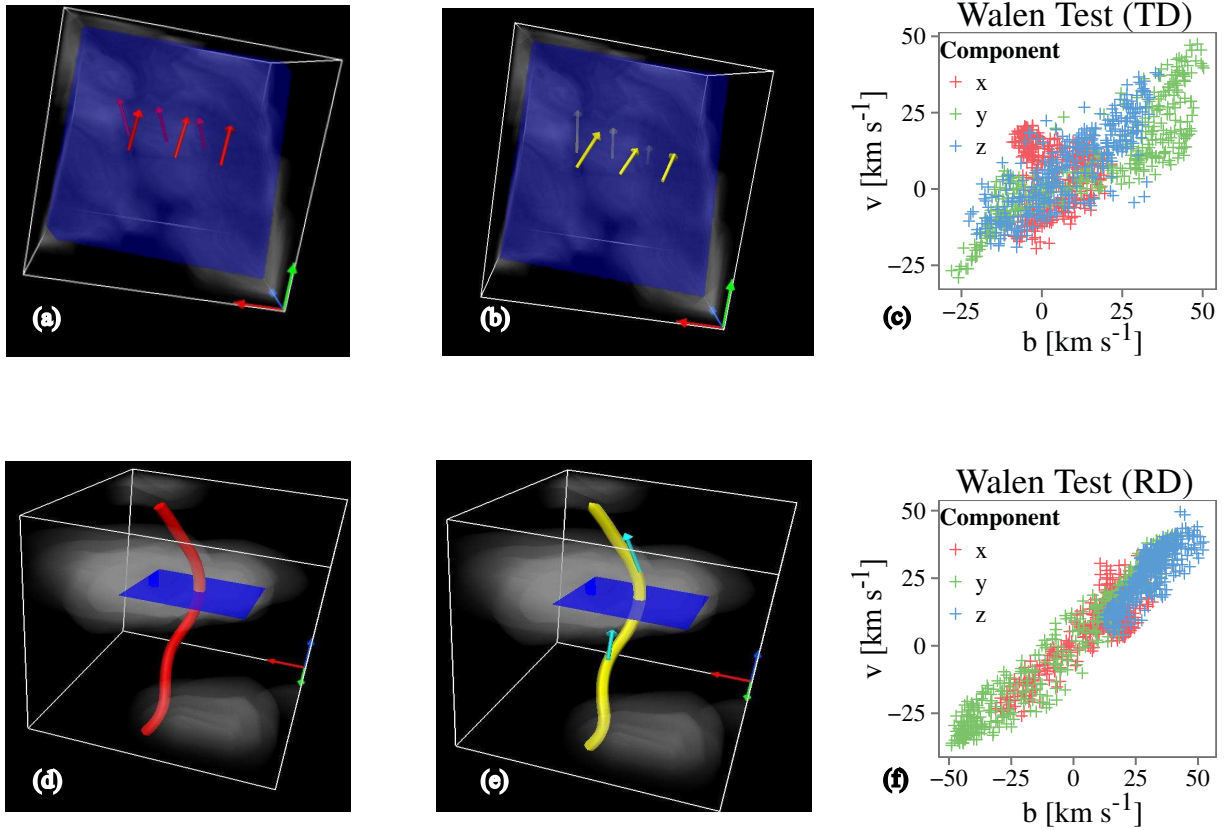}
  \caption{%
  A TD (top panels) and RD (bottom panels) at $t=60\;\mathrm{s}$.
  \textbf{(a) and (b)}: Schematic of the TD, where the transparent blue plane shows
  the plane of discontinuity, the red and yellow arrows denote respectively
  $\mathbf{B}$ and $\mathbf{v}$, the light grey clouds plot the TVI
  with thicker cloud representing larger values, and the red, blue and green
  arrows at the box corner indicate the $x$, $y$ and $z$ directions, respectively.
  The background guide field is along the $z$-direction.
  \textbf{(c)}: Scatter plot of the components of $\mathbf{v}'=\mathbf{v}-\mathbf{v}_{\textrm{HT}}$
  and $\mathbf{b}=\mathbf{B}/\sqrt{\mu_0\rho}$ of the TD.
  \textbf{(d) and (e)}: Schematic of the RD displayed with stream lines;
  cyan arrows show the direction of the velocity in the HT-frame.
  \textbf{(f)}: Scatter plot for the RD.\label{fig2}}
\end{figure}

To check our methods and investigate the physical properties of the
discontinuities, an individual TD and RD are picked and illustrated in
Figure~\ref{fig2}. From the TD data we obtain $|B_n|/|B|_L=0.17$ and
$\delta|B|/|B|_L=0.45$. The TD has a normal 
$\hat{\mathbf n}=(0.893,-0.355,-0.278)$, 
almost aligned to the $x$-axis, and an HT velocity 
$\mathbf{v}_\mathrm{HT}=(4.13,-10.96,-41.99)\;\textrm{km~s}^{-1}$. 
In the HT frame,
Panels (a) and (b) show that $\mathbf{B}$ and $\mathbf{v}$ across the
discontinuity are quasi-parallel to the TD plane. The light grey cloud shows
the sub-volume with high TVI and the plane is soaked therein. Panel (c) gives
the TD's Wal\'en test, where the points are rather scattered, but there still
exists a correlation of $\mathbf{v}$ and $\mathbf{b}=\mathbf{B}/\sqrt{\mu_0
\rho}$, especially in the $y$ and $z$ components (the correlation
coefficients are $r_{xx},r_{yy},r_{zz}=-0.06,0.88,0.82$). For the RD we
find $|B_n|/|B|_L=0.87$, $\delta|B|/|B|_L=0.11$, 
$\hat{\mathbf n}=(0.258,-0.139,0.956)$ and 
$\mathbf{v}_\mathrm{HT}=(1.69,2.17,-55.56)\;\textrm{km s}^{-1}$. Panels (d) and (e) have almost identical and
slightly bent stream lines, thus illustrating the confirmation of the Wal\'en
relation. Panel (f) shows the correlation coefficients 
($r_{xx},r_{yy},r_{zz}=0.87,0.97,0.86$). 
The temperatures at the respective discontinuity points are also computed.
The TD has $T=1.48\times10^5\;\mathrm{K}$, 
$T_\parallel=1.51\times10^5\;\mathrm{K}$, 
and $T_\perp=1.45\times10^5\;\mathrm{K}$, 
while the RD has $T=1.30\times10^5\;\mathrm{K}$, 
$T_\parallel=1.71\times10^5\;\mathrm{K}$, and
$T_\perp=1.09\times 10^5\;\mathrm{K}$. 
The TD is by 13.5\% hotter than the RD in $T$, and by 33.1\% hotter in $T_\perp$.

\begin{figure}
  \plotone{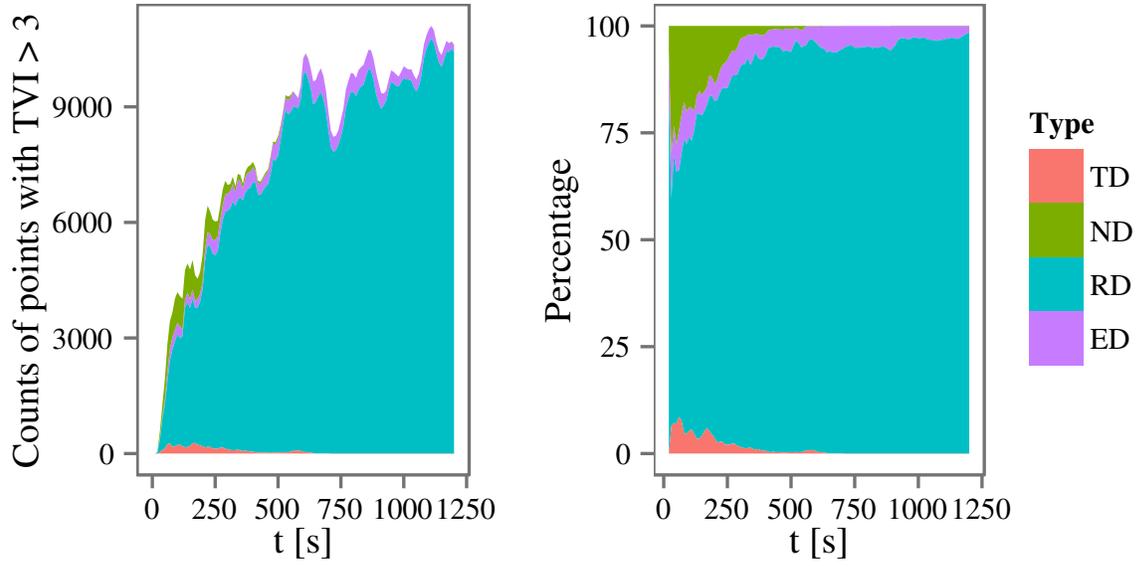}
  \caption{%
  Counts (left) of discontinuity points and their proportions (right) for each type.
  TDs, NDs, RDs and EDs are represented respectively in red, green, cyan and purple.  \label{fig3}} 
\end{figure}

To emphasize furthermore the differences between TDs and RDs, statistical
results are presented in Figure~\ref{fig3}, where we plot the numbers of
discontinuity points of each type as a function of time (left panel), 
as well as their percentages (right panel). 
At $t=0$ there are only a few discontinuity points (53 TDs, 25 RDs, 65 EDs, 65 NDs; RDs are even
not primary), but then the total number of discontinuity points increases
with time, with RD becoming the dominant type. As the temporal evolution
progresses, the number of TDs and their percentage first increase yet
then decrease again slowly, behaving nearly in phase with that
of the changing $\left<j^2\right>$, except during its initial drop phase.

\begin{figure}
  \plotone{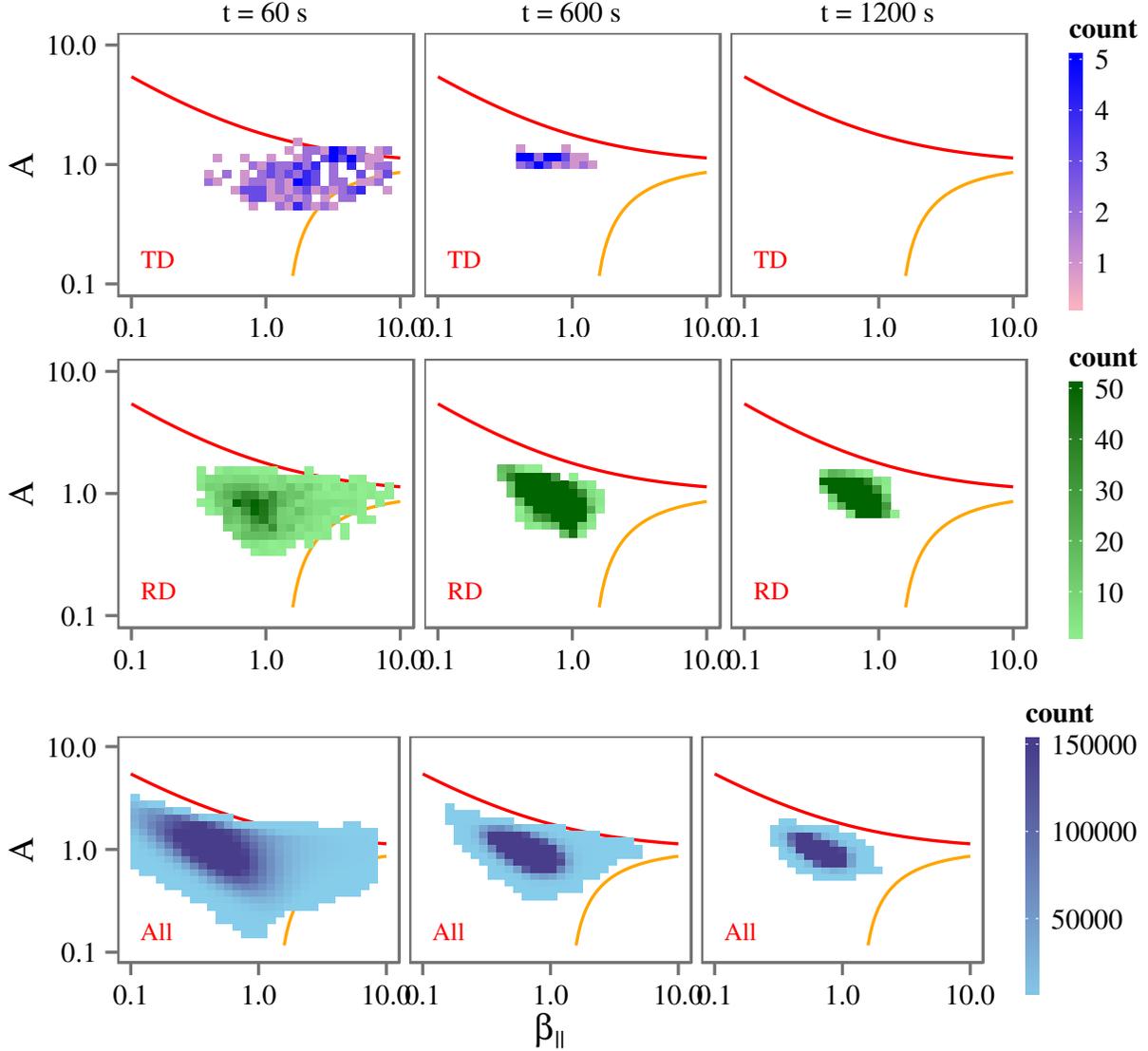}
  \caption{%
  Temporal evolution of the distributions of TD points (top), RD points (centre), 
  \markRevision{and all grid points in the whole computational region (bottom)}
  in the $(\beta_\parallel,A=T_\perp/T_\parallel)$ plane, arranged in columns for different times.
  For reference, the red and orange lines give the thresholds for the mirror and fire-hose instabilities, respectively.
  The colour of a bin denotes the number of grid points therein.  \label{fig4}}
\end{figure}

Moreover, in Figure~\ref{fig4} we plot for TDs and RDs their beta and
anisotropy locations in the $(\beta_\parallel,A)$ plane, where darker bins
correspond to a higher number of discontinuity points, and with a uniform
colour scale to facilitate comparisons for different times. For reference,
the threshold curves of the fire-hose and mirror instabilities
\citep{hellinger06} are also plotted as orange and red curves, respectively.
Apparently, the total numbers of TDs are less than those of RDs. Apart from
that, they also distribute differently. At $t=60\;\mathrm s$, the TD points
tend to aggregate both in the centre region and near to the instability
curves. In the decaying phase, the distribution shrinks toward the centre,
and disappears finally. 
The RD points do not show this trend, with their majority still gathering
around $\beta_\parallel=0.5$, $A=1.0$, i.e.~the initial values. 
At 60~s, some points lie beyond the instability lines but do not congregate there.
\markRevision{For reference, the distributions of all grid points 
are supplied. They resemble those of the RDs. }

\begin{figure}
  \plotone{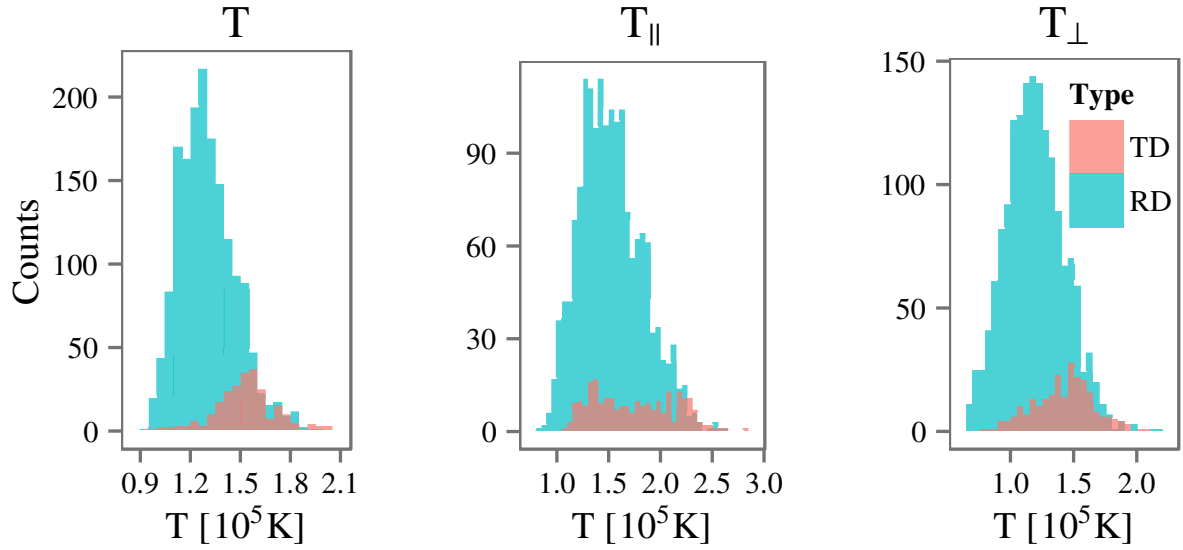}
  \caption{%
  Distributions of the temperatures associated with the TD and RD points.
  The left, middle and right sub-plot shows the distribution of $T$,
  $T_\parallel$ and $T_\perp$, respectively.  \label{fig5}}
\end{figure}

To investigate the heating effect of a discontinuity, the distributions of
the temperatures found at the TD and RD points are also provided. 
Note that simulation of non-adiabatic heating is beyond the scope of this work as no real dissipation is involved. 
In Figure~\ref{fig5} we plot the distributions of $T$, $T_\parallel$ and $T_\perp$ at $t=60\;\mathrm{s}$, with the values
for TDs in red and RDs in blue. The TDs are slightly hotter in $T$ and
$T_\perp$. The TDs have $\bar T=1.53\times 10^5\;\mathrm{K}$ (standard
deviation $\sigma=1.84\times 10^4\;\mathrm{K}$), whereas the RDs have $\bar
T=1.30\times10^5\;\mathrm{K}$ ($\sigma=1.68 \times 10^4\;\mathrm{K}$).
Also, TDs have $\bar T_\perp=1.42\times10^5\;\mathrm{K}$ ($\sigma=2.37\times10^4\;\mathrm{K}$), and RDs have 
$\bar T_\perp=1.19\times10^5\;\mathrm{K}$ ($\sigma=2.31\times10^4\;\mathrm{K}$). Since both types of
discontinuity evolved in the same plasma with a uniform initial temperature,
it is reasonable to conclude that TDs tend to become hotter than RDs, and to
infer that TDs may intrinsically be more heated than RDs. The TDs' increased
temperatures may be caused by plasma squeezing and adiabatic compression. In
the present case study, the TD has a squeezing (quantified as $\nabla_n v_n$)
whose magnitude is 5.41 times that of the RD.

\section{Conclusion}

To conclude, in this letter we describe and apply methods to identify and
analyse the ubiquitous discontinuities appearing in 3D simulation data, and
can therefore present the following statistical results: (1) among the
identified discontinuity points, RDs represent the majority, (2) TDs
aggregate near the extremes of the fire-hose and mirror instability
thresholds, while RDs do not, (3) TDs are hotter than RDs, both in their $T$
and $T_\perp$.

However, this work is also limited in some aspects. Though our simulation of
decaying turbulence in a closed box can reveal its different stages of
evolution in time, the results obtained are difficult to compare with
turbulence observations made in the solar wind, where the convected plasma
volume is open and can always be filled with fresh waves continuously
supplied from a source region. To alleviate this problem (i.e. the difference
between the reality and our case study and its temperature distributions), we
selected the results obtained at the time when $\left<j^2\right>$ is maximal,
which may represent a state being close to developed turbulence.

Moreover, MHD though with thermal anisotropy lacks realistic descriptions of
microscopic (dissipative) solar wind processes.
However, the used MHD model
allows us to reduce significantly the computational cost and to investigate
discontinuities in three-dimensional space, a virtue of our approach which
appears crucial for the analysis. In the identification of discontinuities,
we take $w=3$ and the TVI threshold as 3, values which are reasonably
chosen but seem somehow arbitrary. Therefore, we also checked the outcome
with a different $w$ (ranging from 2 to 5) and TVI threshold (set lower at
2.64) for the statistical study, but found that the results did not change
essentially, in terms of identification, normal direction, proportion, 
and distribution of all discontinuities. 

The physical mechanisms generating TDs and RDs should be further investigated. 
From another aspect, hybrid or full particle simulations should be considered, 
as they enable the description of kinetic processes in the solar wind as well. 
Besides, even in 3D MHD simulations, the methods to
study large TVI events could be further improved, e.g., possibly by
identifying the discontinuities with model structures having a 3D geometric
configuration instead of by simply counting their associated points. Such
approach may improve the identification of the discontinuities and give
better statistics concerning their occurrence rates as well as the
percentages of their local temperature increase.

\acknowledgements

This work is supported by NSFC grants under contracts 41231069, 41174148,
41222032, 41274132, 41474147, 41031066 and 41304133, and was carried out
using the SWMF/BATSRUS tools developed at the University of Michigan Center
for Space Environment Modeling (CSEM) and made available through the NASA
Community Coordinated Modeling Center (CCMC). Figure 2 was partly produced by
VAPOR\citep{vapor}.
JSH, CYT, and XW are also involved in the ISSI/ISSI-BJ international team.

\clearpage

\clearpage



\begin{thebibliography}{}
\bibitem[Bale et al.(2009)]{bale09} Bale, S. D., Kasper, J. C.,
  Howes, G. G., et al., 2009, \prl, 103, 211101
\bibitem[Birn et al.(1995)]{birn95} Birn, J., Gary, S. P., \& Hesse M.
  1995, \jgr, 100, 19211
\bibitem[Borovsky(2008)]{borovsky08} Borovsky, J. E.  2008, \jgr, 113, A08110
\bibitem[Bruno \& Carbone(2013)]{bruno13} Bruno, R., \& Carbone, V.,
  ``The Solar Wind as a Turbulence Laboratory'', LRSP, 10, 2. URL(cited on
  15 Dec 2014):
  http://solarphysics.livingreviews.org/Articles/lrsp-2013-2/
\bibitem[Chew et al(1956)]{cgl} Chew, G.F., Goldberger, M.L., \& Low, F.E. 1956, RSPSA, 236, 112
\bibitem[Clyne et al(2007)]{vapor} Clyne, J., Mininni, P., Norton, A., \&
  Rast, M. 2007, NJPh, 9, 301
\bibitem[Colburn \& Sonett(1966)]{colburn66} Colburn, D. S., \& Sonett, C. P.  1966, \ssr, 5, 439
\bibitem[Goldstein et al.(1995)]{goldstein95} Goldstein, M., Roberts, D.,
  \& Matthaeus, W.  1995, ARA\&A, 33, 283
\bibitem[Greco et al.(2008)]{greco08} Greco, A., Chuychai, P., Matthaeus,
  W. H., Servidio, S., \& Dmitruk, P. 2008, \grl, 35, L19111
\bibitem[Greco et al.(2009)]{greco09} Greco, A., Matthaeus, W. H.,
  Servidio, S., Chuychai, P., \& Dmitruk, P.  2009, \apj, 691, L111
\bibitem[Hellinger et al.(2006)]{hellinger06}Hellinger, P.,
  Tr\'avn\'\i\v cek, P., Kasper, J. C., \& Lazarus, A. J.  2006, \grl, 33,
  L09101
\bibitem[Hesse \& Birn(1992)]{hesse92} Hesse, M., \& Birn, J. 1992, \jgr,
  97, 10643
\bibitem[Karimabadi et al.(2013)]{karimabadi13} Karimabadi, H.,
  Roytershteyn, V., Wan, M. et al., 2013, PhPl, 20, 012303
\bibitem[MacBride et al.(2008)]{macbride08} MacBride, B., Smith, C., \&
  Forman, M.  2008, \apj, 679, 1644
\bibitem[Marsch \& Tu(1994)]{marsch94} Marsch, E., \& Tu, C. Y. 1994, AnGeo,
  12, 1127
\bibitem[Marsch(2006)]{marsch06} Marsch, E., ``Kinetic Physics of the
  Solar Wind'', LRSP, 3, 1. URL(cited on 14 Dec 2014):
  http://solarphysics.livingreviews.org/Articles/lrsp-2006-1/
\bibitem[Matthaeus et al.(1996)]{matthaeus96} Matthaeus, W. H., Ghosh, S.,
  Oughton, S., \& Roberts, D. A.  1996, \jgr, 101, 7619
\bibitem[Meng et al.(2012a)]{meng12} Meng, X., T\'oth, G., Liemohn, M. W.,
  Gombosi, T. I., \& Runov, A.  2012, \jgr, 117, A08216
\bibitem[Meng et al.(2012b)]{meng12b} Meng, X., T\'oth, G., Sokolov, I. V.,
  \& Gombosi, T. I.  2012, JCoPh, 231, 3610
\bibitem[Meng et al.(2013)]{meng13} Meng, X., T\'oth, G., Glocer, A.,
  Fok, M. C., \& Gombosi, T. I.  2013, \jgr, 118, 5639
\bibitem[Miao et al.(2011)]{miao11} Miao, B., Peng, B., \& Li, G.
  2011, AnGeo, 29, 237
\bibitem[Osman et al.(2011)]{osman11} Osman, K. T., Matthaeus, W. H.,
  Greco, A., \& Servidio, S. 2011, \apjl, 727, L11
\bibitem[Osman et al.(2012)]{osman12} Osman, K. T., Matthaeus, W. H.,
  Hnat, B., \& Chapman, S. C. 2012, \prl, 108, 261103
\bibitem[Parashar et al.(2009)]{parashar09} Parashar, T. N., Shay, M. A.,
  Cassak, P. A., \& Matthaeus, W. H.  2009, PhPl, 16, 032310
\bibitem[Paschmann et al.(2013)]{paschmann13} Paschmann, G., Halland, S., 
  Sonnerup, B., \& Knetter, T. 2013, AnGeo, 31, 871
\bibitem[Powell et al.(1999)]{powell99} Powell, K. G., Roe, P. L.,
  Linde, T. J., Gombosi, T. I., \& De Zeeuw, D. L.  1999, JCoPh, 154, 284
\bibitem[Rempel et al.(2009)]{rempel09} Rempel, M., Sch\"ussler, M., \&
  Kn\"olker, M.  2009, \apj, 691, 640
\bibitem[Servidio et al.(2011)]{servidio11} Servidio, S., Dmitruk, P.,
  Greco, A.  et al., 2011, NPGeo, 18, 675
\bibitem[Servidio et al.(2012)]{servidio12} Servidio, S., Valentini, F., 
  Califano, F. \& Veltri, P. 2012, \prl, 108, 045001
\bibitem[Servidio et al.(2014)]{servidio14} Servidio, S., Osman, K. T.,
  Valentini, F.  et al., 2014, \apjl, 781, L27
\bibitem[Smith(1973)]{smith73} Smith, E. J.  1973, \jgr, 78, 2054
\bibitem[Sonnerup \& Cahill(1967)]{sonnerup67} Sonnerup, B. U. \"O. \&
  Cahill, L. J. Jr. 1967, \jgr, 72, 171
\bibitem[Sonnerup et al.(1987)]{sonnerup87} Sonnerup, B. U. \"O.,
  Papamastorakis, I., Paschmann, G., \& L\"uhr, H. 1987, \jgr, 92, 12137
\bibitem[T\'oth et al.(2012)]{toth12} T\'oth, G., van der Holst, B.,
  Sokolov, I. V., et al., 2012, JCoPh, 231, 870
\bibitem[Tu \& Marsch(1995)]{tu95} Tu, C. Y., \& Marsch, E.  1995, \ssr, 73, 1
\bibitem[Wang et al.(2013)]{wang13} Wang, X., Tu, C. Y., He, J. S.,
  Marsch, E., \& Wang, L. H.  2013, \apjl, 772, L14
\end{thebibliography}
\end{document}